 \documentclass[manuscript,screen]{acmart}
\AtBeginDocument{%
  \providecommand\BibTeX{{%
    \normalfont B\kern-0.5em{\scshape i\kern-0.25em b}\kern-0.8em\TeX}}}

\setcopyright{acmlicensed}
\copyrightyear{2018}
\acmYear{2018}
\acmDOI{XXXXXXX.XXXXXXX}

\acmConference[CHI'24 Workshop: LLMs as Research Tools]{}{May 12,
  2024}{Hybrid/Hawaii, US}
\acmBooktitle{CHI '24 Workshop Paper: LLMs as Research Tools: Applications and Evaluations in HCI Data Work, May 12, 2024, Hybrid/Hawaii, United States.} 
\acmISBN{978-1-4503-XXXX-X/18/06}

\begin{document}

\title{The Shiny Scary Future of Automated Research Synthesis in HCI}

\author{Katja Rogers}
\email{katja.rogers@acm.org}
\orcid{0000-0002-5958-3576}
\affiliation{%
  \institution{University of Amsterdam}
  \city{Amsterdam}
  \country{Netherlands}
}

\renewcommand{\shortauthors}{Rogers}

\begin{abstract}
Automation and semi-automation through computational tools like LLMs are also making their way to deployment in research synthesis and secondary research, such as systematic reviews. In some steps of research synthesis, this has the opportunity to provide substantial benefits by saving time that previously was spent on repetitive tasks. The screening stages in particular may benefit from carefully vetted computational support. However, this position paper argues for additional caution when bringing in such tools to the analysis and synthesis phases, where human judgement and expertise should be paramount throughout the process.  
\end{abstract}



\received{04 March 2024}

\maketitle

\section{Introduction}

Automated computational tools (e.g., LLMs and other AI tools) are increasingly being deployed to complete various tasks and workflows in professional settings, including research tasks \cite{eloundou2023gpts, amon2023generative,stahl2023systematic}. This has already led to (sometimes tongue-in-cheek, sometimes earnest) speculation and exploration of whether computational tools could replace humans in the research process altogether \cite{byun2023dispensing,schmidt2024simulating}. 
This is generally motivated through the substantial impact it might be able to have on time saved, especially in the context of secondary research which is generally quite time-consuming~\cite{beller2018making, yao2024evaluating,tsafnat2014systematic}. 

Secondary research, also called research synthesis, is the creation of knowledge from a selection of primary (usually empirical) research papers \cite{gough2004systematic, rogers2023systematic}. For example, forms of secondary research include narrative reviews, systematic reviews, and scoping reviews (although many other types and subtypes exist) \cite{sutton2019meeting}. Typically, secondary research involves stages in which relevant papers are identified and selected to form a corpus of papers for analysis, and then stages in which the papers are analyzed and interpreted to answer a specific question. The details of how both stages are conducted is crucial for the quality of  secondary research. Depending on review type or methodology, they involve critical appraisal and grounded interpretation by experts, and should result in robust answers to specific research questions. The need for careful and rigorous conduct makes the outcome of this kind of research potentially very valuable to the field, but also involves considerable labour and time---which of course, as with any task, people dream of speeding up and making easier.

It is inevitable that computational tools such as large-language models will be used for research synthesis, and it is indeed already happening. Our field needs a timely conversation about how, when, and with which caveats automated computational tools should or can be used for secondary research. In particular, certain stages of research synthesis may be more able to deal with the uncertainty that can arise from computational speed-ups than others; other stages may need to place a premium on comprehensive human involvement and critical nuanced appraisal. This position paper argues that---for systematic reviews, at least---the benefits weigh strongly in the search and selection stage of research, whereas there are a lot more critical and problematic trade-offs to consider in the analysis and synthesis stage.


\section{Background: The Status Quo}
Very broadly, secondary research in "systematic"\footnote{There is no clear definition for this term \cite{martinic2019definition, rogers2023systematic}.} formats like systematic reviews or scoping reviews can be distinguished as having two stages: 1) a \emph{search and selection} stage in which papers are identified to form a relevant corpus, and 2) an \emph{analysis}\footnote{This stage is also termed synthesis, which unfortunately means that research synthesis refers to both the process of secondary research as a whole, and a specific stage within secondary research.} stage in which the corpus papers are (ideally) critically appraised, analyzed, and synthesized to answer a specific research question (or multiple) based on the primary research evidence reported in them. Researchers concerned with this methodology have formulated the so-called "Vienna principles" to guide the automation of systematic reviews~\cite{beller2018making}. 
One of these principles notes that, in general, "Automation may assist with all tasks [in reviews]". However, another emphasizes the need for automation tools to result in "systematic reviews that adhere to high standards".
Previous overviews of automation tools in this space have emphasized that this adherence to expected standards of quality or rigour may be particularly difficult when nuance and critical thinking is required, and/or when subjectivity is inherent to the process \cite{felizardo2020automating,vandinter2021automation, marshall2019toward}. 


\paragraph{Search and Selection}
Across various academic fields, the search and selection phases of reviews are already increasingly drawing on automation tools. 
There are tools that support search (e.g., finding similar papers via Research Rabbit \cite{researchrabbit}, or supporting search query design \cite{wang2022automated}). 
However, by and large computational tools have focused more on supporting screening of papers, i.e., the process through which identified papers are assessed for eligibility (often in two phases, first based on title and abstract, then based on the full text) for being included in a review's analysis. A survey by \citet{scott2021systematic} in the health context showed that this is the stage that researchers are most likely to have used automation tools (79\% of their respondents). In contrast, 15\% had used such to assist in formulating a clear (interventional) research question, and 38\% in the search design or execution.

\citet{bolanos2024artificial} recently identified 19 AI-based tools with varying degrees of computational support for screening in reviews, among them Covidence \cite{covidenceML}, Abstrackr \cite{abstrackr,gates2018technology}, SysRev \cite{bozada2021sysrev}, and Iris.ai \cite{irisai}. Similarly, \citet{khalil2022tools} conducted scoping review of automation tools in this context, and declared that "Abstract screening has reached maturity", though they also noted many reported limitations---including a lack of generalizability (e.g., tools tested only on when searching for specific, very targeted studies like randomized controlled trials), and that many tools still need to be validated in comparison to human reviewers without automation.


One such example of computationally supported screening is ASReview \cite{vandeschoot2021open}, which targets the screening stage specifically. Researchers conducting a review can input their list of identified potentially relevant papers, and then flag a number of them (e.g., 20) as relevant or irrelevant. ASReview then uses this as its training to re-order the stack of papers, from most to least likely in relevance. The manual screening process then begins with the papers most likely to be relevant, and can be terminated after a pre-determined stopping criterion of a specific number of irrelevant papers (e.g., 100), as the rest of the stack should be even less relevant. 

ASReview has been shown to reduce time spent on screening significantly in a simulation study \cite{ferdinands2020active}, though the literature does not yet show a test of its accuracy/validity against human reviewers. 
\citet{vandijk2023artificial} reported on using it in a review, stating it "considerably reduced the number of articles in the screening process [...] only 23\% of the total number of articles were screened before the predefined stopping criterion was met. Assuming that all relevant articles were found, the AI tool saved 77\% of the time for title and abstract screening."

In an overview by \citet{wagner2022artificial}, the search and selection phases of reviews have been described as holding "very high [...and] high potential" for benefits from computational tool integration, respectively. They reason that these phases include repetitive and laborious tasks that lend themselves well to automation, though they also note that this potential becomes more moderate for the second full-text screening "which requires considerable expert judgment (especially for borderline cases)" \cite{wagner2022artificial}.

\paragraph{Research Synthesis}
This part of the review stage involves---depending on one's definitions---quality assessment, data extraction, and analysis of the identified corpus of relevant papers, as well as the synthesis and interpretation of results across that corpus. 
\citet{wagner2022artificial} rate the potential for automation to be mostly low to moderate in terms of full automation. They rate the potential high for formal data extraction of simple descriptive data (e.g., sample size in a study), and even descriptive syntheses via text mining---though the potential of more interpretive approaches remains a stark "very low". 
The survey by \citet{scott2021systematic} among clinical researchers reported that 51\% of respondents had used automation tools for data extraction (including risk of bias data), and 46\% for data synthesis/meta-analysis, although it is not clearly reported which tools performed these tasks, or which specific steps were automated.

\section{Computational Automation in Human-Computer Interaction Research Synthesis}

So what does this mean for research synthesis in human-computer interaction (HCI)? 

\paragraph{Shiny Future}
The benefits are certainly enticing; in the future we will be able to conduct reviews much more readily than today. Computational tools like LLMs will be able to assist us in formulating a research question, determining whether it has been answered already, developing and refining the search strategy, and gathering the initial set of relevant papers. It may take on the bulk of screening to arrive at the review corpus, output descriptive summaries of coded characteristics based on text mining and large-language models, and may even eventually run specific analyses on the data for us. Perhaps LLMs (or similar tools) will even suggest some interpretations of the findings for us to review, before we instruct it to write up the reporting of the review according to a specific set of guidelines. Finally, it is even possible it would produce an interactive website with information visualization to display the findings as a living review \cite{elliott2017living}. 
Essentially, we as researchers only need design the review plan (with assistance), then it is completed for us to only double-check, ``\emph{at the push of a button}''~\cite{tsafnat2013automation}. 
With this, we as researchers will benefit from significant time savings \cite{yao2024evaluating} and will be able to put that new free time to good use to write more papers, fix the peer review system, or simply to get more rest.



In line with \citet{wagner2022artificial}, there seem clear potential benefits to the use of automation tools in the search and screening phases, as long as they remain accompanied by human judgment to ensure that the tools are set up to conduct steps within appropriate parameters, to step in when screening eligibility decisions are ambiguous or uncertain, and to double-check results. 
It is unclear how well underway the future described above is for HCI, but the present is certainly already moving in that direction for screening: one review I was involved in already employed ASReview  \cite{wijkstra2023help}, and it did indeed save time---with a pre-defined stopping criteria of 10\% of the initial dataset being labelled irrelevant in a row, ASReview allowed us to terminate screening at \textasciitilde42\% of the total initial set of identified papers. 

For search, automation may also be agreeable in HCI. Partly this is because our reviews already have to contend with the fact that our digital libraries do not make it easy to create reproducible searches, especially across databases \cite{rogers2023systematic}. Further, compared to other fields, our reviews often do not aim to gather all papers on a topic anyway, instead opting to survey specific publication venues as a representative sample \cite{hornbaek2017technology}, or a sample from a specific year \cite{caine2016local} or year range \cite{sabie2022decade, zhou2021dance}; and we almost never include gray literature. Given such common approaches to value representativeness over comprehensiveness, why indeed not use computational tools to speed up the process?

\paragraph{Scary Future}
Alternatively, or possibly in parallel to the above, computational automation of research synthesis will supercharge the speed at which papers are produced---and especially the speed at which surface-level, primarily descriptive secondary research is produced.

In HCI, the synthesis stage is already often not described in much detail. Looking at reviews in HCI, for the most part we do not actually do "proper" systematic reviews in of the sense of its origins in the medical field; focused interventional research questions (what is the effect of X on Y) and meta-analyses are rare. Instead, we more commonly conduct reviews to answer broad research questions (often in the style of scoping reviews \cite{arksey2005scoping} or systematic mapping studies \cite{petersen2015guidelines}). Our community appears to view reporting quality of reviews as relegated to whether there is a (PRISMA \cite{page2021prisma}) figure to illustrate the search and selection process \cite{stefanidi2023literature}; we do not have clear guidance for synthesis steps. 

With this as our status quo, throwing computational tools into the mix for the research synthesis stage appears reckless. Analysis and synthesis always also requires a descriptive understanding of corpus papers, However, for many review types and methods, large crucial parts of synthesis require interpretation, nuance, and contextual understanding---exactly the parts that computational approaches still struggle with. 
In an era of automation-driven research, our research community may be tempted to opt for the kind of secondary research that is easy for computational tools to perform for us, i.e., surface-level scoping reviews that summarize easily-coded characteristics. There is nothing wrong with such contributions and they can indeed be very useful, but in a healthy field, this should not be the only kind of secondary research that is done. 
A strong focus on specific review methods that are conductive to AI support may also lead our research community to disavow more complex methodologies---which are likely required given the complexity and diversity of primary research in HCI \cite{rogers2023systematic}.



\section{Conclusion}
Like in many other ongoing discussions on computational AI tools, quality research synthesis will require that we keep human expertise and critical thinking involved at all key stages of the process. The  use of computational tools needs to be evaluated carefully, but especially so for the synthesis step. Quoting \citet{yao2024evaluating}, "Despite the promise these tools show, caution is advised due to limited evidence. Until further advancements and comprehensive evaluations are undertaken, AI tools should serve as a complement rather than a complete replacement to human reviewers."
As we increasingly embrace these tools, we have to also vet their accuracy, as well as learn to embrace uncertainty and be transparent when it comes to the subjectivity involved in the process---with and without the involvement of computational tools.



\bibliographystyle{ACM-Reference-Format}
\bibliography{★BIBLIO}

\end{document}